\documentclass[onecolumn,showpacs,11pt]{revtex4}
\usepackage{graphicx}
\usepackage{dcolumn}
\usepackage{bm}
\usepackage{epstopdf}
\begin{document}
\newcommand{\hs}{\hspace*{0.3cm}}
\newcommand{\vs}{\vspace*{0.3cm}}
\newcommand{\be}{\begin{equation}}
\newcommand{\ee}{\end{equation}}
\newcommand{\bea}{\begin{eqnarray}}
\newcommand{\eea}{\end{eqnarray}}
\newcommand{\ben}{\begin{enumerate}}
\newcommand{\een}{\end{enumerate}}
\newcommand{\bde}{\begin{widetext}}
\newcommand{\ede}{\end{widetext}}
\newcommand{\nn}{\nonumber}
\newcommand{\crn}{\nonumber \\}
\newcommand{\Tr}{\mathrm{Tr}}
\newcommand{\non}{\nonumber}
\newcommand{\noi}{\noindent}
\newcommand{\al}{\alpha}
\newcommand{\la}{\lambda}
\newcommand{\bet}{\beta}
\newcommand{\ga}{\gamma}
\newcommand{\va}{\varphi}
\newcommand{\om}{\omega}
\newcommand{\pa}{\partial}
\newcommand{\+}{\dagger}
\newcommand{\fr}{\frac}
\newcommand{\bc}{\begin{center}}
\newcommand{\ec}{\end{center}}
\newcommand{\Ga}{\Gamma}
\newcommand{\de}{\delta}
\newcommand{\De}{\Delta}
\newcommand{\ep}{\epsilon}
\newcommand{\varep}{\varepsilon}
\newcommand{\ka}{\kappa}
\newcommand{\La}{\Lambda}
\newcommand{\si}{\sigma}
\newcommand{\Si}{\Sigma}
\newcommand{\ta}{\tau}
\newcommand{\up}{\upsilon}
\newcommand{\Up}{\Upsilon}
\newcommand{\ze}{\zeta}
\newcommand{\ps}{\psi}
\newcommand{\Ps}{\Psi}
\newcommand{\ph}{\phi}
\newcommand{\vph}{\varphi}
\newcommand{\Ph}{\Phi}
\newcommand{\Om}{\Omega}

\title{$T_7$ flavor symmetry scheme for understanding neutrino mass and
mixing in 3-3-1 model with neutral leptons}

\author{V. V. Vien}
\email{wvienk16@gmail.com}
\affiliation{Department of Physics, Tay Nguyen University, 567 Le
Duan, Buon Ma Thuot, DakLak, Vietnam}

\date{\today}

\begin{abstract}
We construct a new version for the 3-3-1 model based on $T_7$
flavor symmetry where the left-handed leptons under $T_7$ differ
from those of our previous work while the $\mathrm{SU}(3)_C
\otimes \mathrm{SU}(3)_L \otimes \mathrm{U}(1)_X$ gauge symmetry
is retain. The flavor mixing patterns and mass splitting are
obtained without perturbation. The realistic lepton mixing can be
obtained if both the direction of breakings $T_7 \rightarrow Z_3$
and $Z_3 \rightarrow \{\mathrm{Identity}\}$ are taken place in
neutrino sector. Maximal CP violation is predicted and CKM matrix
is the identity matrix at the tree-level.
\end{abstract}


 \maketitle

\section{\label{intro} Introduction}
 Despite the great success of the Standard Model (SM)
of the elementary particle physics, the origin of flavor
structure, masses and mixings between generations of matter
particles are unknown yet. The neutrino mass and mixing is one of
the most important evidence of beyond Standard Model physics and
also one of the biggest challenges in particle physics. Many
experiments show that neutrinos have tiny masses and their mixing
is sill mysterious \cite{altar1,altar2, add1,add2,add3,add4,add5}.
The tri-bimaximal form for explaining the lepton mixing scheme was
first proposed by Harrison-Perkins-Scott (HPS), which apart from
the phase redefinitions, is given by \cite{hps1, hps2, hps3, hps4}
\begin{eqnarray}
U_{\mathrm{HPS}}=\left(
\begin{array}{ccc}
\frac{2}{\sqrt{6}}       &\frac{1}{\sqrt{3}}  &0\\
-\frac{1}{\sqrt{6}}      &\frac{1}{\sqrt{3}}  &\frac{1}{\sqrt{2}}\\
-\frac{1}{\sqrt{6}}      &\frac{1}{\sqrt{3}}  &-\frac{1}{\sqrt{2}}
\end{array}\right),\label{Uhps}
\end{eqnarray}
which can be considered  as a good approximation for the recent
neutrino experimental data. The best fit values of neutrino mass
squared differences and the leptonic mixing angles in
Refs.\cite{Nuexp1, Nuexp2} have been given in Tables \ref{NormalH}
and \ref{InvertedH}.
\begin{table}
\caption{\label{NormalH} The experimental values of neutrino mass
squared splittings and leptonic mixing parameters, taken from Refs.
\cite{Nuexp1,Nuexp2} for normal hierarchy.}
\begin{center}
\begin{tabular}{lllllll}
\hline\noalign{\smallskip}
Parameter & Best fit & $1\sigma $ range& $2\sigma $ range \\
\noalign{\smallskip}\hline\noalign{\smallskip}
$\Delta m_{21}^{2}$($10^{-5}$eV$^2$) & $7.62$ & $7.43-7.81$ & $7.27-8.01$ \\

$\Delta m_{31}^{2}$($10^{-3}$eV$^2$)& $2.55$ & $2.64-2.61$ & $2.38-2.68$ \\

$\sin ^{2}\theta _{12}$ &  $0.320$ & $0.303-0.336$ & $0.29-0.35$   \\
 $\sin ^{2}\theta _{23}$&  $0.613$ & $0.573-0.635$ & $0.38-0.66$ \\

 $\sin ^{2}\theta_{13}$& $0.0246$ & $0.0218-0.0275$ & $0.019-0.03$  \\
\noalign{\smallskip}\hline
\end{tabular}
\end{center}
\end{table}
\begin{table}
\caption{\label{InvertedH} The experimental values of neutrino mass
squared splittings and leptonic mixing parameters, taken from Refs.
\cite{Nuexp1, Nuexp2} for inverted hierarchy.}
\begin{center}
\begin{tabular}{lllllll}
\hline\noalign{\smallskip}
Parameter & Best fit & $1\sigma $ range& $2\sigma $ range \\
\noalign{\smallskip}\hline\noalign{\smallskip}
$\Delta m_{21}^{2}$($10^{-5}$eV$^2$) & $7.62$ & $7.43-7.81$ & $7.27-8.01$ \\

$\Delta m_{31}^{2}$($10^{-3}$eV$^2$)& $2.43$ & $2.37-2.50$ & $2.29-2.58$ \\

$\sin ^{2}\theta _{12}$ &  $0.32$ & $0.303-0.336$ & $0.29-0.35$  \\
 $\sin ^{2}\theta _{23}$&  $0.60$ & $0.569-0.626$ & $0.39-0.65$ \\

 $\sin ^{2}\theta_{13}$& $0.025$ & $0.0223-0.0276$ & $0.02-0.03$\\
\noalign{\smallskip}\hline
\end{tabular}
\end{center}
\end{table}
These large neutrino mixing angles are completely different from the quark
 mixing ones defined by the Cabibbo- Kobayashi-Maskawa (CKM) matrix \cite{CKM}.
 This has stimulated work on flavor symmetries and non-Abelian discrete
 symmetries, which are considered to be the most attractive candidate to
formulate dynamical principles that can lead to the flavor mixing
patterns for quarks and lepton. There are many recent models based
on the non-Abelian discrete symmetries, for example, see Ref.\cite{vlT7} and references there in.

In the SM, CP symmetry is violated due
to a complex phase in the CKM matrix \cite{CKM, CKM1}. However,
since the extent
 of CP violation in the SM is not enough for achieving the observed BAU,  we need new source of CP violation for successful BAU.
On the other hand, CP violations in the lepton sector are
imperative if the BAU could be realized through leptogenesis. So,
any hint or observation of the leptonic CP violation  can
strengthen our belief in leptogenesis \cite{CPvio}. The violation of the CP symmetry is a crucial ingredient of any
dynamical mechanism which intends to explain both low energy CP
violation and the baryon asymmetry. Renormalizable gauge theories
are based
 on the spontaneous symmetry breaking mechanism, and it is natural to have the spontaneous CP violation
  as an integral part of that mechanism. Determining all
possible sources of CP violation is a fundamental challenge for high energy physics. In theoretical
 and economical viewpoints, the spontaneous CP breaking necessary to generate the baryon asymmetry
  and leptonic CP violation at low energies brings us to a common source which comes from the
  phase of the scalar field responsible for the spontaneous CP breaking at a high energy scale \cite{CPvio}.

Among the standard model's extensions, the 3-3-1 models have
interesting features which have been introduced in
Refs.\cite{dlshA4,dlsvS4,dlnvS3,vlS4,vlS3,vlD4,vlT7,vD4}. In Refs.
\cite{dlshA4,dlsvS4} we have studied the 3-3-1 model with neutral
fermions based
    on $A_4$ and $S_4$ groups, in which the exact tribimaximal form is obtained,
     where $\theta_{13}= 0$. As we know, the recent considerations have implied $\theta_{13}\neq 0$ as shown in Tables \ref{NormalH}, \ref{InvertedH}. This problem has been improved in the 3-3-1 model with other
      non-Abelian discrete symmetries \cite{dlnvS3,vlS4,vlS3,vlD4,vlT7,vD4}. In Ref. \cite{vlT7} we have studied the 3-3-1 model
with neutral fermions based
    on $T_7$ group, in which all three left-handed fermion fields are set in one triplet $\underline{3}$ under $T_7$.
    In this paper, we investigate another choice for this type of 3-3-1 model based on $T_7$ discrete symmetry in
    which three left-handed lepton fields are put in the $\underline{3}^*$ instead of $\underline{3}$ under $T_7$, and the 3 generations
of right-handed lepton singlets are put, respectively, in the
$\underline{1}$, $\underline{1}''$ and $\underline{1}'$ instead of
$\underline{1}$, $\underline{1}'$ and $\underline{1}'$ of $T_7$.
The motivation for the change to our previous work studied in
Ref.\cite{vlT7} is derived from a useful feature of tensor
products of $T_7$ group. Namely, $3\otimes 3\otimes 3$ or
$3^*\otimes 3^*\otimes 3^*$ has two invariants and $3\otimes
3\otimes 3^*$ or $3^*\otimes 3^*\otimes 3$ has one invariant.
Hence, we can propose another choice of fermion content of the
model to obtain desired results.
    The motivation for extending the above application to the 3-3-1
models with the neutral fermions $N_R$ is mentioned in
\cite{dlshA4, dlsvS4, dlnvS3}.

The rest of this work is organized
as follows. In Sec. \ref{fermion} and \ref{Chargedlep} we present
the necessary elements of the model as well as introducing necessary Higgs fields responsible for the
charged lepton masses. Sec. \ref{neutrino} is devoted for the neutrino mass and
mixing. We summarize our results and make conclusions in the
section \ref{conclus}.

\section{\label{fermion} Fermion content}

The fermion content of the model is similar to that in \cite{vlT7} except three left-handed fermions with $T_7$ symmetry. Under the $[\mathrm{SU}(3)_L, \mathrm{U}(1)_X,
\mathrm{U}(1)_\mathcal{L},\underline{T}_7]$ symmetries as
proposed, the fermions of the model transform as follows \bea
\psi_{L} &\equiv& \psi_{1,2,3L}=\left( \nu_{L} \hs
    l_{L} \hs
    N^c_{R}\right)^T\sim [3,-1/3,2/3,\underline{3}^*],\crn
l_{1R}&\sim&[1,-1,1,\underline{1}],\hs
l_{2 R}\sim[1,-1,1,\underline{1}''], \hs
l_{3 R}\sim[1,-1,1,\underline{1}'].\eea where the subscript numbers
on field indicate to respective families which also  in order
define components of their $T_7$ multiplets. In the following, we
consider possibilities of generating the masses for the fermions.
The scalar multiplets needed for the purpose are also introduced.

\section{\label{Chargedlep} Charged lepton masses}
    The charged lepton masses arise from the couplings of $\bar{\psi}_{L} l_{1R}, \bar{\psi}_{L} l_{2R}$
     and $\bar{\psi}_{L} l_{3R}$ to scalars, where $\bar{\psi}_{L} l_{iL}\, (i=1,2,3)$ transforms as $3^*$ under
$\mathrm{SU}(3)_L$ and $\underline{3}^*$  under
$T_7$. To generate masses for charged leptons, we need a $SU(3)_L$ Higgs triplets that lying
in $\underline{3}$ under $T_7$ and transforms as $3$ under $\mathrm{SU}(3)_L$,
\bea \phi_i = \left(%
\begin{array}{c}
  \phi^+_{i1} \\
  \phi^0_{i2} \\
  \phi^+_{i3} \\
\end{array}%
\right)\sim [3,2/3,-1/3, \underline{3}^*] \hs (i=1,2,3) \label{phi}.\eea
In this work, we
argue that $T_7\rightarrow Z_3$ in charged - lepton sector is taken place, and this can be achieved by the Higgs triplet $\phi$ with the VEV alignment $\langle \phi\rangle=(\langle \phi_1\rangle, \langle \phi_1\rangle, \langle \phi_1\rangle)$ under $T_7$\cite{vlT7}, where
\be \langle \phi_1 \rangle = \left(0 \hs v \hs 0 \right)^T.\label{vevphi} \ee
The Yukawa interactions are
 \bea -\mathcal{L}_{l}&=&h_1 (\bar{\psi}_{L}\phi)_{\underline{1}} l_{1R}+h_2 (\bar{\psi}_{L}\phi)_{\underline{1}'}l_{2R}
+h_3 (\bar{\psi}_{i L}\phi)_{\underline{1}''} l_{3R}+H.c\crn
&=&h_1
(\bar{\psi}_{1L}\phi_1+\bar{\psi}_{2L}\phi_2+\bar{\psi}_{3L}\phi_3)l_{1R}\crn
&+&h_2 (\bar{\psi}_{1L}\phi_1+\om\bar{\psi}_{2L}\phi_2+\om^2
\bar{\psi}_{3L}\phi_3)l_{2R}\crn &+&h_3
(\bar{\psi}_{1L}\phi_1+\om^2\bar{\psi}_{2L}\phi_2+\om
\bar{\psi}_{3L}\phi_3) l_{3R}+H.c.\eea
The mass Lagrangian for the charged leptons is then given by
\bea
-\mathcal{L}^{\mathrm{mass}}_l&=&h_1 v\bar{l}_{1L} l_{1R}+h_2 v\bar{l}_{1L} l_{2R}+h_3 v\bar{l}_{1L} l_{3R}\crn
&+&h_1v\bar{l}_{2L} l_{1R}+h_2v \om\bar{l}_{2L} l_{2R}+h_3v \om^2\bar{l}_{2L} l_{3R}\crn
&+&h_1v\bar{l}_{3L} l_{1R}+h_2v\om^2 \bar{l}_{3L} l_{2R}+h_3v\om \bar{l}_{3L} l_{3R}+H.c. \eea
The mass
Lagrangian for the charged leptons reads
\bea
-\mathcal{L}^{\mathrm{mass}}_l=(\bar{l}_{1L},\bar{l}_{2L},\bar{l}_{3L})
M_l (l_{1R},l_{2R},l_{3R})^T+H.c, \eea
where \be M_l=
\left(%
\begin{array}{ccc}
  h_1v&\,\,\, h_2v &\,\,\, h_3 v \\
   h_1v & \,\,\,\,\,\,\, \om h_2 v &\,\,\,\, \om^2 h_3 v  \\
  h_1v & \,\,\,\,\, \om^2 h_2 v  &\,\,\,\,\,\,\om h_3 v \\
\end{array}%
\right).\label{Mltq}\ee
This matrix can be diagonalized as,
\bea U^{\dagger}_L M_lU_R=\left(%
\begin{array}{ccc}
  \sqrt{3}h_1 v & 0 & 0 \\
  0 & \sqrt{3}h_2 v & 0 \\
  0 & 0 & \sqrt{3}h_3 v \\
\end{array}%
\right)\equiv \left(%
\begin{array}{ccc}
  m_e & 0 & 0 \\
  0 & m_\mu & 0 \\
  0 & 0 & m_\tau \\
\end{array}%
\right),\label{Mld}\eea where \bea U_L=\fr{1}{\sqrt{3}}\left(%
\begin{array}{ccc}
  1 &\,\,\, 1 &\,\,\, 1 \\
  1 &\,\,\, \om &\,\,\, \om^2 \\
  1 &\,\,\, \om^2 &\,\,\, \om \\
\end{array}%
\right),\hs U_R=1.\label{Uclep}\eea
We note that the charged - lepton mixing matrix in (\ref{Uclep}) is different to that of in Ref. \cite{vlT7}. This leads to the difference of the lepton mixing matrix of these two versions, and this is the main result of this work.

The experimental  values for masses of  the charged leptons at the weak
scale are given as \cite{PDG2012}  : \bea m_e=0.511\, \textrm{MeV},\hs \
 m_{\mu}=105.658 \ \textrm{MeV},\hs m_{\tau}=1776.82\,
\textrm{MeV} \eea from which it follows that $h_1\ll h_2\ll h_3$.
On the other hand, if we choose the VEV $v\sim 100 GeV$ then $h_1
\sim 10^{-6},\, h_2\sim 10^{-4},\, h_3\sim 10^{-2}$.

In similarity to the charged lepton sector, to generate the quark
masses, we  additionally introduce the two scalar Higgs triplets $
\eta, \chi$ respectively lying in $\underline{3}$ and
$\underline{1}$ under $T_7$. By assuming that the VEVs
    of $\eta$ and $\chi$ are given as $\langle\eta\rangle=
(\langle\eta_1\rangle, \langle\eta_1\rangle,\langle\eta_1\rangle)$
with $\langle\eta_1\rangle=\left( u \hs   0 \hs 0\right)^T$ and $
\langle\chi\rangle=\left( 0 \hs   0 \hs   v_\chi\right)^T$, from
the invariant Yukawa interactions, the exotic quarks therefore get
masses \cite{vlT7}: $m_U=f_3 v_\chi ,\hs m_{D_{1,2}}=f_{1,2}
v_\chi$, and the masses of ordinary up-quarks and down-quarks are
\bea m_u &=&-\sqrt{3}h^u_1 v,\hs m_c=-\sqrt{3}h^u_2 v,\hs m_t
=\sqrt{3}h^u_3 u,\crn m_d&=&\sqrt{3}h^d_1 u,\hs\,\,\,\,\,
m_s=\sqrt{3}h^d_2 u,\hs\,\,\,\,\, m_b =\sqrt{3}h^d_3 v. \eea
  The unitary matrices which couple the left-handed quarks $u_L$ and
$d_L$ to those in the mass bases are unit ones. The CKM quark
mixing matrix at the tree level is then
$U_\mathrm{CKM}=U^{\dagger}_{dL} U_{uL}=1$. This is a good
approximation for the realistic quark mixing matrix, which implies
that the mixings among the quarks are small. To obtain a realistic
quark spectrum, we should add radiative correction or use the
effective six-dimensional operators (see Ref. \cite{car} for
details). However, we leave this problem for the future work.
 A detailed study on charged lepton and quark masses can be found in
 Ref. \cite{dlnvS3}. In this paper, I consider a new version for the 3-3-1 model based on $T_7$
flavor symmetry responsible for neutrino mass and mixing.

\section{\label{neutrino} Neutrino mass and mixing}
The neutrino masses arise from the couplings of $\bar{\psi}^c_{L} \psi_{L}$ to scalars,
 where $\bar{\psi}^c_{L} \psi_{L}$ transforms as $3^*\oplus 6$ under
$\mathrm{SU}(3)_L$ and $\underline{3}^*\oplus \underline{3}\oplus
\underline{3}$  under $T_7$. To obtain a realistic neutrino spectrum, the antisextets transform
as follows
 \bea \sigma_i&=&
\left(%
\begin{array}{ccc}
  \sigma^0_{11} & \sigma^+_{12} & \sigma^0_{13} \\
  \sigma^+_{12} & \sigma^{++}_{22} & \sigma^+_{23} \\
  \sigma^0_{13} & \sigma^+_{23} & \sigma^0_{33} \\
\end{array}%
\right)_i\sim [6^*,2/3,-4/3,\underline{3}],\label{sigma}\\
s_i &=&
\left(%
\begin{array}{ccc}
  s^0_{11} & s^+_{12} & s^0_{13} \\
  s^+_{12} & s^{++}_{22} & s^+_{23} \\
  s^0_{13} & s^+_{23} & s^0_{33} \\
\end{array}%
\right)_i \sim [6^*,2/3,-4/3,\underline{3}^*], \label{s}\\
 \sigma'_i&=&
\left(%
\begin{array}{ccc}
  \sigma'^0_{11} & \sigma'^+_{12} & \sigma'^0_{13} \\
  \sigma'^+_{12} & \sigma'^{++}_{22} & \sigma'^+_{23} \\
  \sigma'^0_{13} & \sigma'^+_{23} & \sigma'^0_{33} \\
\end{array}%
\right)_i\sim [6^*,2/3,-4/3,\underline{3}], \hs (i=1,2,3) \label{sip}\eea
The alignments of anti-sextets were explained in Ref.\cite{vlT7}. In this work we also
argue that both the breakings $T_7\rightarrow Z_3$ and
$T_7\rightarrow \{\mathrm{identity}\}$ (Instead of $Z_3\rightarrow \{\mathrm{identity}\}$)
 must be taken place in neutrino sector. However, the VEVs of $\si$
does only one of these tasks. These happen with the VEVs alignments as follows \cite{vlT7}:
\bea
\langle \si\rangle&=&(\langle
 \si_1\rangle, \langle \si_1\rangle, \langle \si_1\rangle),\hs \langle \si_1\rangle=\left(%
\begin{array}{ccc}
  \la_{\si} & 0 & v_{\si} \\
  0 & 0 & 0 \\
  v_{\si} & 0 & \La_{\si} \\
\end{array}%
\right),\label{sivev}\\
\langle s \rangle &=& (\langle s_1 \rangle, 0,0)^T,\hs
\langle s_1\rangle=\left(%
\begin{array}{ccc}
  \la_{s} & 0 & v_{s} \\
  0 & 0 & 0 \\
  v_{s} & 0 & \La_{s} \\
\end{array}%
\right), \label{svev}\\
\langle \si' \rangle &=& (\langle \si'_1 \rangle, 0,0)^T,\hs \langle \si'_1\rangle =\left(%
\begin{array}{ccc}
  \la'_{\si} & 0 & v'_{\si} \\
  0 & 0 & 0 \\
  v'_{\si} & 0 & \La'_{\si} \\
\end{array}%
\right).
\label{sipvev} \eea
 Note that the alignments of anti-sextets $\si, s,\si'$ are the same as those in Ref.\cite{vlT7}, the unique difference is under $T_7$ representations.
 The Yukawa interactions in the neutrino sector are:
 \bea -\mathcal{L}_\nu&=&\fr 1 2 x
(\bar{\psi}^c_L \si)_{{\underline{3}}^*}\psi_L+y (\bar{\psi}^c_L
s)_{{\underline{3}}^*}\psi_L +\fr z 2(\bar{\psi}^c_L
\si')_{{\underline{3}}^*}\psi_L+H.c.\crn &=& \fr 1 2 x
(\bar{\psi}^c_{1L}\si_2\psi_{1L}+\bar{\psi}^c_{2L}\si_3
\psi_{2L}+\bar{\psi}^c_{3L}\si_1\psi_{3L})\crn &+&
y(\bar{\psi}^c_{2L}s_3\psi_{1L}+\bar{\psi}^c_{3L}s_1
\psi_{2L}+\bar{\psi}^c_{1L}s_2\psi_{3L})\crn &+&\fr z 2
(\bar{\psi}^c_{1L}\si'_2\psi_{1L}+\bar{\psi}^c_{2L}\si'_3
\psi_{2L}+\bar{\psi}^c_{3L}\si'_1\psi_{3L}) +H.c.\label{yn}\eea
Although the Yukawa interactions in Eq.(\ref{yn}) are differ from
the one give in Ref.\cite{vlT7}, the mass terms for the neutrino
sector are the same, which is given by \bea
-\mathcal{L}^{\mathrm{mass}}_\nu &=&\fr 1 2
x(\la_{\si}\bar{\nu}^c_{1 L}\nu_{1L}+ v_{\si}\bar{\nu}^c_{1
L}N^c_{1R}+v_{\si}\bar{N}_{1R}\nu_{1L}+\La_{\si}\bar{N}_{1R}N^c_{1R})\crn
&+&\fr 1 2 x(\la_{\si}\bar{\nu}^c_{2
L}\nu_{2L}+v_{\si}\bar{\nu}^c_{2L}N^c_{2R}+
v_{\si}\bar{N}_{2R}\nu_{2L}+\La_{\si}\bar{N}_{2R}N^c_{2R})\crn &+&
\fr 1 2 x (\la_\si\bar{\nu}^c_{3 L}\nu_{3L}+v_\si\bar{\nu}^c_{3
L}N^c_{3R}+
v_\si\bar{N}_{3R}\nu_{3L}+\La_\si\bar{N}_{3R}N^c_{3R})\crn
&+&y(\la_{s}\bar{\nu}^c_{3 L}\nu_{2L}+v_{s}\bar{\nu}^c_{3
L}N^c_{2R}+
v_{s}\bar{N}_{3R}\nu_{2L}+\La_{s}\bar{N}_{3R}N^c_{2R})\crn &+&\fr
1 2 z(\la'_\si\bar{\nu}^c_{3 L}\nu_{3L}+v'_\si\bar{\nu}^c_{3
L}N^c_{3R}+
v'_\si\bar{N}_{3R}\nu_{3L}+\La'_\si\bar{N}_{3R}N^c_{3R}) +H.c.
\label{T7nm}\eea Therefore, neutrino mass and mixing are the same
as those in Ref.\cite{vlT7}. Namely, \bea m_1 &=&\fr 1 2 \left(B_1
+ B_2 + \sqrt{(B_1 - B_2)^2+4C^2}\right),\crn
m_2&=&A,\label{m123}\\
m_3&=&\fr 1 2 \left(B_1 + B_2 - \sqrt{(B_1 - B_2)^2+4C^2}\right),\nn\eea
and \bea U_\nu=\left(%
\begin{array}{ccc}
  0 & 1 & 0 \\
  \fr{1}{\sqrt{K^2+1}} & 0 & \fr{K}{\sqrt{K^2+1}} \\
  -\fr{K}{\sqrt{K^2+1}} & 0 & \fr{1}{\sqrt{K^2+1}} \\
\end{array}%
\right).\left(%
\begin{array}{ccc}
  1 & 0 & 0 \\
 0 & 1 & 0 \\
 0 & 0& i \\
\end{array}%
\right),\label{Unu1}\eea
where
\bea
A&=& a_L-\frac{a^2_D}{a_R},\crn
B_1&=&a_L-\frac{a_Rb^2_D-2a_Db_Db_R+a^2_D(a_R+d_R)}{a^2_R-b^2_R+a_Rd_R}\crn
B_2&=&B_1+d_L+\frac{2(b_Db_R-a_Da_R)d_D+(a^2_D-b^2_D)d_R-a_Rd^2_D}{a^2_R-b^2_R+a_Rd_R},\crn
C&=&b_L-\frac{(a^2_D+b^2_D)b_R-(2a_Da_R + a_Dd_R)b_D+(a_Db_R -a_Rb_D)d_D}{a^2_R-b^2_R+a_Rd_R},\label{ABC}
\eea
and
\bea
K&=&\frac{B_1 -B_2 -\sqrt{4 C^2 + (B_1-B_2)^2}}{2C}.\label{K}
\eea
with $a_{L,D,R}, b_{L,D,R}$ and $c_{L,D,R}$ are given in Eq. (5.15) in Ref.\cite{vlT7}. \\
Combining (\ref{Uclep}) and (\ref{Unu1}), we get the lepton mixing matrix:
\bea U_{lep}=U^\dagger_L
U_\nu= \fr{1}{\sqrt{3}}\left(%
\begin{array}{ccc}
  \fr{1-K}{\sqrt{K^2+1}} & 1 &  \fr{1+K}{\sqrt{K^2+1}} \\
 \fr{\om(\om-K)}{\sqrt{K^2+1}} & 1 &  \fr{\om(K\om+1)}{\sqrt{K^2+1}} \\
  \fr{\om(1-K\om)}{\sqrt{K^2+1}}  & 1 &  \fr{\om(\om+K)}{\sqrt{K^2+1}} \\
\end{array}%
\right).\left(%
\begin{array}{ccc}
  1 & 0 & 0 \\
 0 & 1 & 0 \\
 0 & 0& i \\
\end{array}%
\right).\label{Ulep}\eea It is worth noting that in this version
the lepton mixing matrix in (\ref{Ulep}) takes a similar form but
not identical to that in Ref.\cite{vlT7}. The unique difference is
the role of $K$ in the lepton mixing matrix $U_{lep}$. On the
other hand, the matrix $U_{lep}$ given in (\ref{Ulep}) is slightly
different from $U_{HPS}$ in (\ref{Uhps}), but
 similar to the original version of trimaximal mixing considered
in Ref. \cite{TBM2, MaximalCP} which is based on the $\Delta(27)$ group extension of the Standard Model. Although there are some
phenomenological predictions of the model are similar to those in Ref.  \cite{TBM2, MaximalCP} but the fundamental difference between our model with the well known one is the prediction of CP violation. Namely, our model predicts maximal CP violation $\delta =\pi/2$ with $\theta_{23}\neq \pi/4$ whereas in Ref. \cite{TBM2}, the maximal CP violation $\delta =\pi/2, 3\pi/2$ achieved with $\theta_{23}= \pi/4$. \\
In the standard Particle Data Group (PDG) parametrization, the lepton mixing
 matrix ($U_{PMNS}$) can be parametrized as
 \be
       U_{PMNS} = \left(%
\begin{array}{ccc}
    c_{12} c_{13}     & -s_{12} c_{13}                    & -s_{13} e^{-i \delta}\\
    s_{12} c_{23}-c_{12} s_{23} s_{13}e^{i \delta} & c_{12} c_{23}+s_{12} s_{23} s_{13} e^{i \delta} &-s_{23} c_{13}\\
    s_{12} s_{23}+c_{12} c_{23} s_{13}e^{i \delta}&c_{12} s_{23}-s_{12} c_{23} s_{13} e^{i \delta}  & c_{23} c_{13} \\
    \end{array}%
\right) \times P, \label{Ulepg}
\ee
where $P=\mathrm{diag}(1, e^{i \alpha}, e^{i \beta})$, and
$c_{ij}=\cos \theta_{ij}$, $s_{ij}=\sin \theta_{ij}$ with
$\theta_{12}$, $\theta_{23}$ and $\theta_{13}$ being the solar
angle, atmospheric angle and the reactor angle respectively.
$\delta$ is the Dirac CP violating phase while $\alpha$ and
$\beta$ are the two Majorana CP violating phases.

By comparing Eqs. (\ref{Ulep}) and (\ref{Ulepg}) we obtain
$\alpha=0, \beta =\frac{\pi}{2}$ for the two Majorana phases, and
the lepton mixing matrix in (\ref{Ulep}) can be parameterized in
terms of three Euler's angles $\theta_{ij}$ as follows:
 \bea s_{13} e^{-i \delta}&=&\fr{-1-K}{\sqrt{3}\sqrt{K^2+1}},\label{s13}\\
 t_{12}&=&\frac{\sqrt{K^2+1}}{K-1},\label{t12}\\
  t_{23}&=&-\frac{1+K\om}{\om+K}.\label{t23}
\eea Substituting $\om =-\frac{1}{2}+i\frac{\sqrt{3}}{2}$ into
(\ref{t23}) yields: \bea K&=&k_1+ik_2,\crn
k_1&=&\frac{1}{2}\frac{t^2_{23}-4t_{23}+1}{t^2_{23}-t_{23}+1},\hs
k_2=\frac{\sqrt{3}}{2}\frac{1-t^2_{23}}{t^2_{23}-t_{23}+1}.\label{K12}
\eea The expression (\ref{K12})  tells us that $k^2_1+k^2_2\equiv
|K|^2=1$. Combining (\ref{s13}) and (\ref{t12}) yields: \bea
e^{-i\delta}&=&\frac{1}{\sqrt{3}s_{13}t_{12}}\frac{1+K}{1-K}
=\frac{i}{s_{13}t_{12}}\frac{1-t_{23}}{1+t_{23}}
\equiv\cos\delta-i\sin\delta \nn\eea or \bea \cos\delta&=&0,\hs
\sin\delta
=\frac{t_{23}-1}{s_{13}t_{12}(t_{23}+1)}.\label{sindel} \eea
Since $\cos\delta=0$ so that $\sin\delta$ must be equal to $\pm1$,
it is then $\delta =\frac{\pi}{2}$ or $\delta =\frac{3\pi}{2}$. However, to fit the data in Tab. \ref{NormalH} and Tab. \ref{InvertedH}, $\delta =\frac{\pi}{2}$ is used. Thus, this version predicts the maximal Dirac CP violating phase which
  is the same as our previous work and also similar to the results in Refs. \cite{TBM2, MaximalCP}. We emphasize that the maximal CP
  violation $\delta =\frac{\pi}{2}$ in our model achieved with $\theta_{23}\neq \pi/4$ in contrast to that in Refs.\cite{TBM2, MaximalCP}, and this is one of the most striking prediction
  of the model under consideration.
  In the case $\delta = \frac{\pi}{2}$, from (\ref{sindel}) we have the relation
 among three Euler's angles as follows: \bea
t_{23}&=&\frac{1+s_{13}t_{12}}{1-s_{13}t_{12}}
=\frac{\sqrt{1-\sin^2\theta_{12}}+\sqrt{\sin^2\theta_{12}\sin^2\theta_{13}}}{
\sqrt{1-\sin^2\theta_{12}}-\sqrt{\sin^2\theta_{12}\sin^2\theta_{13}}}.
\label{relat1} \eea
\begin{itemize}
\item[(i)]
For the best fit values of $\theta_{12}$ and $\theta_{13}$,in the normal case, given in Table \ref{NormalH}, $\sin^2\theta_{12}=0.320 $ and $\sin^2\theta_{13}=0.0246$, we obtain
 $t_{23}=1.24113 \, (\theta_{23}=51.141^o)$, and \be
K=-0.932872 - 0.360207 i, \,\, (|K|=1). \label{Kvalues}\ee
The lepton mixing matrix in (\ref{Ulep}) then takes the form:
\be U\simeq\left(%
\begin{array}{ccc}
 0.831055 &\hs 0.57735 &\hs 0.154874 \\
-0.549652 &\hs 0.57735 &\hs -0.797152 \\
-0.281403 &\hs 0.57735 &\hs  0.642278 \\
\end{array}%
\right).\label{Ulepmix1}\ee  The value of the Jarlskog invariant
$J_{CP}$ which determines the magnitude of CP violation in
neutrino oscillations is determined \cite{Jarlskog}: \bea
J_{CP} =\frac{1}{8}\cos\theta_{13}\sin2\theta_{12}\sin2\theta_{23}\sin2
\theta_{13}\sin\delta
= 0.034865.\label{J1}
\eea
\item[(ii)]
In a similarity way, for the best fit values of $\theta_{12}$ and $\theta_{13}$,in the inverted case, given in Table \ref{InvertedH}, $\sin^2\theta_{12}=0.320 $ and $\sin^2\theta_{13}=0.0250$, we obtain
 $t_{23}=1.24332 \, (\theta_{23}=51.190^o)$, and \be
K=-0.931818 - 0.362925 i, \,\, (|K|=1). \label{Kvalues1}\ee
The lepton mixing matrix in (\ref{Ulep}) then takes the form:
\be U\simeq\left(%
\begin{array}{ccc}
 0.831298 &\hs 0.57735 &\hs 0.156174 \\
-0.5509 &\hs 0.57735 &\hs -0.798012 \\
-0.2804 &\hs 0.57735 &\hs  0.641839 \\
\end{array}%
\right).\label{Ulepmix1}\ee  The value of the Jarlskog invariant
$J_{CP}$ is determined \cite{Jarlskog}: \bea
J_{CP} =0.03512.\label{J1}
\eea
\end{itemize}
Up to now the values of neutrino masses (or the absolute neutrino
masses) as well as the mass ordering of neutrinos is unknown.
The neutrino mass spectrum can be the normal mass hierarchy ($
|m_1|\simeq |m_2| < |m_3|$), the inverted hierarchy ($|m_3|< |m_1|\simeq |m_2|$)
 or nearly degenerate ($|m_1|\simeq |m_2|\simeq |m_3| $). The mass
ordering of neutrino depends on the sign of $\Delta m^2_{23}$
which is currently unknown. 
From (\ref{m123}),(\ref{Kvalues}) or(\ref{Kvalues1}) and the two
experimental constraints on squared mass differences of neutrinos
as shown in Tab.\ref{NormalH} and Tab.\ref{InvertedH}, we have the
solutions as shown below.

\subsection{Normal case ($\Delta m^2_{23}> 0$)}
In this case, combining (\ref{m123}), (\ref{K}) with the two
experimental constraints on squared mass differences of neutrinos
as shown in Tab. \ref{NormalH}, we get a solution (in [eV]) as
follows: \bea C&=&0.5\sqrt{\ep_1-2\sqrt{\ep_2}},\hs
B_1=B_2+(6.98504\times 10^{-7}-0.720414 i)C,\crn B_2&=&-0.5\sqrt{4
A^2-0.0003048} -(3.49252\times 10^{-7}-0.360207i)C\crn
&-&0.5\sqrt{(3.481-1.00642\times 10^{-6}i)C^2},\crn
m_1&=&-0.5\sqrt{4 A^2-0.0003048},\,\,\, m_2=A,\crn
m_3&=&-0.5\sqrt{4 A^2-0.0003048}-\sqrt{\Ga_{-}} \label{case1} \eea
where \bea \ep_1 &=&0.00275507+7.96543\times 10^{-10}i +
(2.29819+6.6445\times 10^{-7}i)A^2,\crn \ep_2 &=&-2.48903\times
10^{-7}-1.43925\times 10^{-13}i +(0.00316583+1.83061\times
10^{-9}i)A^2\crn &+&(1.32042+7.63516\times 10^{-7}i)A^4,\crn
\Ga_{-}&=&0.0023976-1.03398\times 10^{-25}i +(2-1.05879\times
10^{-22}i)A^2\crn &-&(1.7405-5.03212\times
10^{-7}i)\sqrt{\ep_2}.\label{Gam} \eea

\begin{figure}[ht]
\bc
\includegraphics[width=7.0cm, height=5.5 cm]{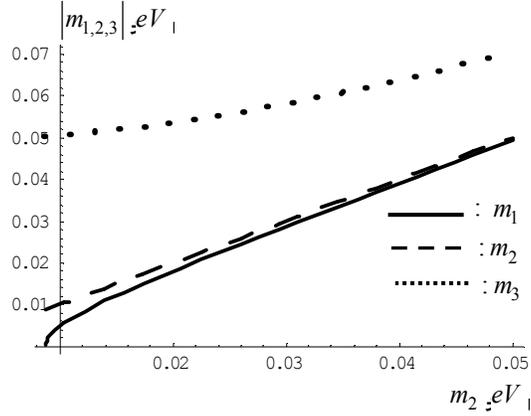}
\vspace*{-0.1cm} \caption[$|m_{1,3}|$ as functions of $m_2$ with
 $m_2\in(0.00867, 0.05) \, \mathrm{eV}$ in
 the case of $\Delta m^2_{23}> 0$.]{$|m_{1,3}|$ as functions of $m_2$ with
 $m_2\in(0.00867, 0.05) \, \mathrm{eV}$ in
 the case of $\Delta m^2_{23}> 0$.}\label{m123N01T7}
\ec
\end{figure}

\begin{figure}[ht]
\bc
\includegraphics[width=7.0cm, height=5.5 cm]{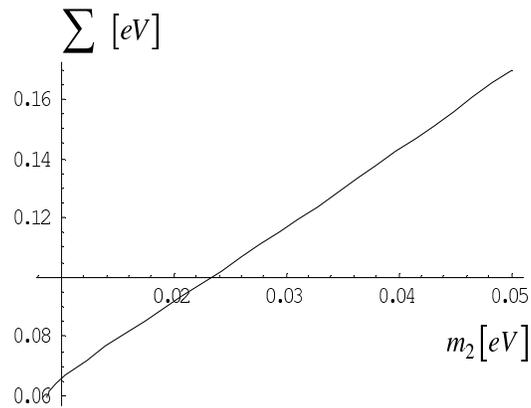}
\vspace*{-0.1cm} \caption[The sum $\sum=\sum^3_{i=1}|m_i|$ as a function of
$m_2$ with $m_2 \in (0.00867, 0.05)\,\mathrm{eV}$ in the
case of $\Delta m^2_{23}> 0$]{The sum $\sum=\sum^3_{i=1}|m_i|$ as a function of
$m_2$ with $m_2 \in (0.00867, 0.05)\,\mathrm{eV}$ in the
case of $\Delta m^2_{23}> 0$.}\label{m123Ns01T7}
\ec
\end{figure}

In Fig. \ref{m123N01T7} we have plotted the absolute values $|m_{1,3}|$
as functions of $m_2$ with  $m_2 \in (0.00867, 0.05)\, \mathrm{eV}$. This figure shows
  that there exist allowed regions for values $m_2$ (or $A$) where either normal
  or quasi-degenerate neutrino masses spectrum is achieved.
  The quasi-degenerate mass hierarchy is obtained when $|A|$ lies
   in a region [$0.05\,\mathrm{eV} , +\infty$]  ($|A|$ increases
  but must be small enough because of the scale of $m_{1,2,3}$). The normal
  mass hierarchy will be obtained if $|A|$ takes the values around $(0.0087, 0.01)\,
   \mathrm{eV}$.
The Fig. \ref{m123Ns01T7} gives the sum
   $\sum=\sum^3_{i=1}|m_i|$ with $m_2 \in (0.0087, 0.05)\,\mathrm{eV}$.

\begin{figure}[ht]
\begin{center}
\includegraphics[width=7.0cm, height=5.5 cm]{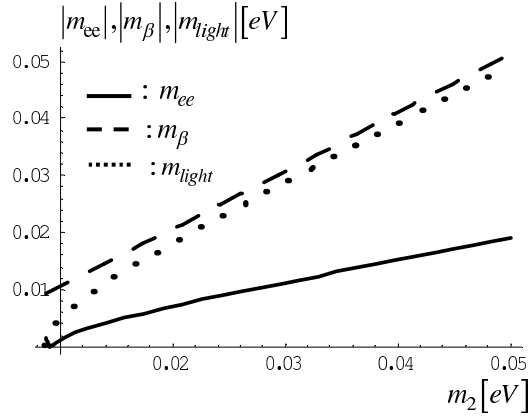}
\vspace*{-0.1cm} \caption[$|m_{ee}|$, $|m_{\beta}|$ and $|m_{light}|$ as functions of $m_2$
from (\ref{case1}) in the case of $\Delta m^2_{23}> 0$ with $m_2\in(0.00867, 0.05) \,
\mathrm{eV}$.]{$|m_{ee}|$, $|m_{\beta}|$ and $|m_{light}|$ as functions of $m_2$
from (\ref{case1}) in the case of $\Delta m^2_{23}> 0$ with $m_2\in(0.00867, 0.05) \,
\mathrm{eV}$.}\label{mee.T7}
\end{center}
\end{figure}

From the expressions (\ref{m123}), (\ref{Ulep}), it is easily
 to obtain the effective mass $\langle m_{ee}\rangle$ governing neutrinoless double beta decay
 \cite{betdecay1, betdecay2,betdecay3,betdecay4,betdecay5, betdecay6},$\langle m_{ee}\rangle = \mid\sum^3_{i=1} U_{ei}^2 m_i \mid $ and $m_\beta = \sqrt{\sum^3_{i=1} |U_{ei}|^2 m_i^2 }$ which have been studied in detailed in \cite{vlT7}. In the normal spectrum, $|m_1|\approx |m_2|<|m_3|$, so $m_1\equiv m_{light}$ is the lightest neutrino mass. In Fig. \ref{mee.T7} we have plotted the value $|m_{ee}|$,
$|m_{\beta}|$ and $|m_{light}|$ as functions of $m_2$ with $m_2 \in (0.0087, 0.05)\,\mathrm{eV}$.
To get explicit values of the model
parameters, we assume $m_2=10^{-2}\, \mathrm{eV}$, which is safely small.
Then the other neutrino masses are explicitly
 given as $m_1\simeq-5.298\times 10^{-3}\, \mathrm{eV},\, m_2\simeq 10^{-2}\, \mathrm{eV},\,
  m_3\simeq -4.95 \times 10^{-2} \, \mathrm{eV}$ and $|m_{ee}|\simeq 1.09
  \times 10^{-3} \, \mathrm{eV},\, |m_{\beta}|\simeq 1.178 \times 10^{-2} \, \mathrm{eV}$.

\subsection{Inverted case ($\Delta m^2_{23}< 0$)}
\begin{figure}[ht]
\begin{center}
\includegraphics[width=7.0cm, height=5.5cm]{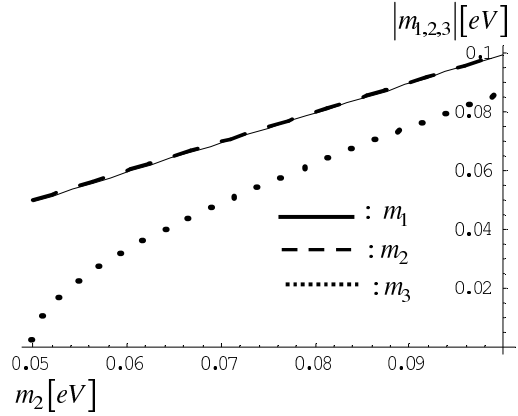}
\vspace*{-0.5cm} \caption[$|m_{1,3}|$ as functions of
$m_2$  in the
case of $\Delta m^2_{23}< 0$ with $m_2\in (0.05, 0.1) \, \mathrm{eV}$]{$|m_{1,3}|$ as functions of
$m_2$  in the
case of $\Delta m^2_{23}< 0$ with $m_2\in (0.05, 0.1) \, \mathrm{eV}$.}\label{m123I01T7}
\end{center}
\end{figure}

\begin{figure}[ht]
\begin{center}
\includegraphics[width=7.0cm, height=5.5cm]{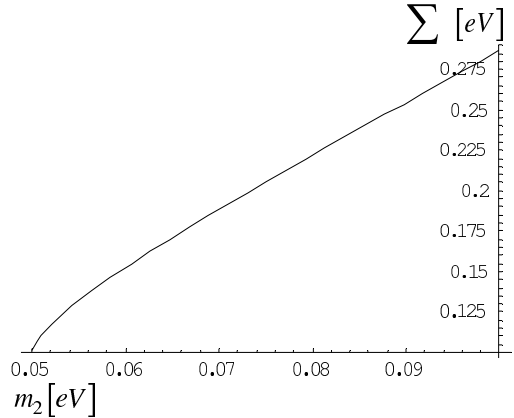}
\caption[The sum $\sum$ as a function of $m_2$ with $m_2\in (0.05, 0.1)\,
\mathrm{eV}$ in the case of $\Delta m^2_{23}<
0$]{The sum $\sum$ as a function of $m_2$ with $m_2\in (0.05, 0.1)\,
\mathrm{eV}$ in the case of $\Delta m^2_{23}<
0$.}\label{m123Is01T7}
\end{center}
\end{figure}

\begin{figure}[ht]
\begin{center}
\includegraphics[width=6.5cm, height=5.0cm]{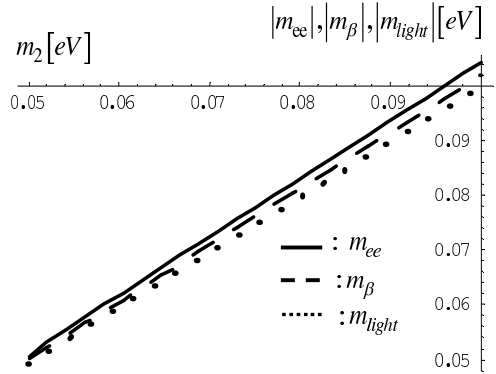}
\vspace*{-0.1cm} \caption[$|m_{ee}|$, $|m_{\beta}|$ and $|m_{light}|$ as functions of $m_2$ in the case of $\Delta m^2_{23}< 0$ with $m_2\in
(0.05, 0.1) \, \mathrm{eV}$.]{$|m_{ee}|$, $|m_{\beta}|$ and $|m_{light}|$ as functions of $m_2$ in the case of $\Delta m^2_{23}< 0$ with $m_2\in
(0.05, 0.1) \, \mathrm{eV}$.}\label{meeI.T7}
\end{center}
\end{figure}

Similar to the normal case, in this case we also have a solution as follows:
 \bea
C&=&0.5\sqrt{\al_1-2\sqrt{\beta_1}},\hs B_1=B_2+(6.98504\times 10^{-7}-0.720414 i)C,\crn
B_2&=&0.5\sqrt{4
A^2-0.0003048}-(3.49252\times 10^{-7}-0.360207i)C\crn
&-&0.5\sqrt{(3.481-1.00642\times 10^{-6}i)C^2},\crn
m_1&=&0.5\sqrt{4 A^2-0.0003048},\,\,\, m_2=A,\crn m_3&=&0.5\sqrt{4
A^2-0.0003048}-\sqrt{\ga_{-}}, \label{case9} \eea
where
\bea \al_1
&=&(-0.00296742-8.57938\times 10^{-10}i)+
(2.29819+6.6445\times 10^{-7}i)A^2,\crn
\beta_1&=&2.52163\times
10^{-7}+1.4581\times 10^{-13}i-(0.00340984+1.9717\times
10^{-9}i)A^2\crn &+&(1.32042+7.63516\times 10^{-7}i)A^4,\crn
\ga_{-}&=&-0.0025824+1.03398\times 10^{-25}i
+(2-1.05879\times 10^{-22}i)A^2\crn &\mp&(1.7405-5.03212\times
10^{-7}i)\sqrt{\beta_1}.\label{gam} \eea
 The absolute value $|m_{1,3}|$ as functions of $m_2$ with $m_2\in
(0.05, 0.1) \, \mathrm{eV}$ is plotted in Fig. \ref{m123I01T7}.
This figure shows that there exist
allowed regions for value of $m_2$ (or $A$) where either inverted or quasi-degenerate neutrino mass hierarchy achieved. The quasi-degenerate
  mass hierarchy obtained when $|A|$ lies in a region [$0.1\,\mathrm{eV} , +\infty$]($|A|$ increases but must be small enough because of
  the scale of $m_{1,2,3}$). The
 inverted mass hierarchy will be obtained if $A$ takes the values
  around $(0.05, 0.1)\, \mathrm{eV}$.
Fig. \ref{m123Is01T7} gives the sum $\sum$ with $m_2 \in (0.05, 0.1) \,\mathrm{eV}$.
In the inverted spectrum, $|m_3|< |m_1|\simeq |m_2|$, and $m_3 \equiv m_{light}$ is the lightest neutrino mass. $|m_{ee}|$,
$|m_{\beta}|$ and $|m_{light}|$ as functions of $m_2$ with $m_2
\in (0.05, 0.1)\,\mathrm{eV}$ is plotted in Fig. \ref{meeI.T7}.

 In similarity to the normal case, to get explicit values of the
model parameters, we assume $m_2=5\times10^{-2}\, \mathrm{eV}$,  which is safely small.
 Then the other neutrino masses are explicitly
 given as $ m_1 \simeq 4.925\times 10^{-2}\, \mathrm{eV}$ and $m_3\simeq 1.342\times
 10^{-2} \, \mathrm{eV}$.

\section{\label{conclus}Conclusions}
In this paper, we have modified the previous 3-3-1 model combined
with $T_7$ discrete symmetry to adapt recent neutrino mixing with
non-zero $\theta_{13}$. We have shown that the realistic neutrino
masses and mixings can be obtained if the two directions of
breakings $T_7 \rightarrow Z_3$ and $Z_3 \rightarrow
\{\mathrm{Identity}\}$ are taken place in neutrino sector and are
equivalent in size, i.e, the contributions due to $s$, $\si$ and
$s'$ are comparable. The model predicts maximal CP violation with
$\theta_{23}\neq \frac{\pi}{4}$.
\section*{Acknowledgments}
This work was supported in part by the National Foundation for
Science and Technology Development (NAFOSTED) of Vietnam under
Grant No: 103.01-2014.51.

\end{document}